\documentclass[twocolumn,prb,showpacs]{revtex4}
\usepackage{graphicx}
\usepackage{dcolumn}
\usepackage{bm}
\usepackage{amsmath}
\usepackage{color}
\usepackage{times}
\usepackage{soul}

\begin{document}

\preprint{}
\title{Josephson effect through magnetic skyrmion}
\author{Takehito Yokoyama$^1$ and Jacob Linder$^2$}
\affiliation{$^1$Department of Physics, Tokyo Institute of Technology, Tokyo 152-8551,
Japan 
\\ $^2$Department of Physics, Norwegian University of Science and Technology, N-7491 Trondheim, Norway
}
\date{\today}

\begin{abstract}
We discover that the multiple degrees of freedom associated with magnetic skyrmions: size, position, and helicity, can all be used to control the Josephson effect and 0-$\pi$ transitions occurring in superconductor/magnetic skyrmion/superconductor junctions. In the presence of two skyrmions, the Josephson effect depends strongly on their relative helicity and leads to the possibility of a helicity-transistor effect for the supercurrent where the critical current is changed by several orders of magnitude simply by reversing the helicity of a magnetic skyrmion. 
Moreover, we demonstrate that the Fraunhofer pattern can show a local minimum at zero flux as a direct result of the skyrmion magnetic texture.
These findings demonstrate the rich physics that emerges when combining topological magnetic objects with superconductors and could lead to new perspectives in superconducting spintronics.
\end{abstract}

\maketitle

The interplay between superconductivity and ferromagnetism in hybrid structures has received much attention in recent years \cite{Buzdinrev,Bergeretrev}, due to its allure from a fundamental physics viewpoint and also because of improved and new functionality brought about by using superconductors in spintronics \cite{linder_nphys_15}. Due to the proximity effect, the Cooper pairs induced in the ferromagnet acquire a finite center of mass momentum. Therefore, the pair amplitude oscillates in space which may result in a sign change of the Josephson current in ferromagnetic Josephson junctions. This effect can be used to control the quantum ground state of the system, altering from a state where the superconducting phase difference is 0 to a state where it is $\pi$ \cite{Bulaevskii,Buzdin,Buzdin2,Golubov}. This 0-$\pi$ transition was originally observed in Josephson junctions through weak ferromagnets. \cite{Ryazanov,Kontos} Also, in the presence of inhomogeneity of the magnetic order, triplet pairing with spin aligned with the local exchange field is generated in the ferromagnet due to spin flip scattering \cite{eschrig_07,Bergeret3}. Experiments have successfully demonstrated the presence of such spin-triplet pairing by observing a Josephson current through strong ferromagnets \cite{Keizer,Khaire,Robinson}, which can be explained via the concepts of spin-mixing and spin-rotation taking place near the superconductor/ferromagnet interface \cite{Eschrig}. Using equal spin triplet pairings, the possibility arises to enhance existing effects or discover new ones in spintronics \cite{linder_nphys_15, giazotto_prb_08, romeo_prl_13, trif_prl_13, machon_prl_13, linder_prb_14}.
Recently, it has been also proposed that inhomogeneous ferromagnet/superconductor junctions can create topological superconductivity. \cite{Martin,Lu,Nakosai}

Currently, much interest is garnered by magnetic skyrmions in chiral magnets \cite{Rossler,Muhlbauer,Nagaosa2}. Such objects are characterized by a topologically protected spin configuration. Due to their peculiar magnetic structure, several intriguing phenomena have been discovered such as topological and skyrmion Hall effects \cite{Lee,Neubauer,Zang} and current-driven motion of skyrmion with ultralow current density \cite{Jonietz,Fert,Iwasaki,Iwasaki2,Troncoso}.
It has been shown that magnetic skyrmions can be also driven by a temperature gradient \cite{Kong,Mochizuki,Lin,Kovalev}.
A thermal gradient is predicted to induce a skyrmion motion towards the high temperature region accompanied by a skyrmion Hall effect \cite{Kong}. Skyrmions are accompanied by a degree of freedom known as their \textit{helicity}, which is determined by their spin swirling direction. It has been experimentally demonstrated that the helicity of skyrmions can be changed both via a small external magnetic field \cite{yu_pnas_12} and spin-orbit interactions \cite{shibata_natnan_13}. This opens the exciting prospect that any physical quantity that responds to a change in the skyrmion helicity degree of freedom will be controllable via an external field.

In this paper, we investigate how the degrees of freedom of magnetic skyrmions, such as helicity, can influence the supercurrent-response and quantum ground state of Josephson junctions including skyrmions. We find that the supercurrent is strongly influenced by the (i) size, (ii) position, and (iii) helicity of magnetic skyrmions. We discover that the 0-$\pi$ transition can in fact be triggered by changing any of these three skyrmion properties, which in turn have been confirmed to be experimentally tunable via external magnetic fields \cite{yu_pnas_12}, spin-orbit interactions \cite{shibata_natnan_13}, and electric currents \cite{Jonietz, yu_ncom_12}. This offers a new and dynamical way of manipulating the quantum ground state of a superconducting system via magnetic skyrmions. We then show that the strong dependence on the helicity creates a \textit{helicity-transistor} effect for supercurrents, where the critical current is changed by several orders of magnitude upon reversing the helicity of a skyrmion. Moreover, we find that as a direct consequence of the skyrmion magnetization texture, the Fraunhofer pattern can display a local minimum at zero flux in contrast to conventional homogeneous magnetic Josephson junctions.
 In what follows, we will demonstrate these properties in Josephson junctions featuring both single and two skyrmions.

\begin{figure}[tbp]
\begin{center}
\scalebox{0.8}{
\includegraphics[width=11.0cm,clip]{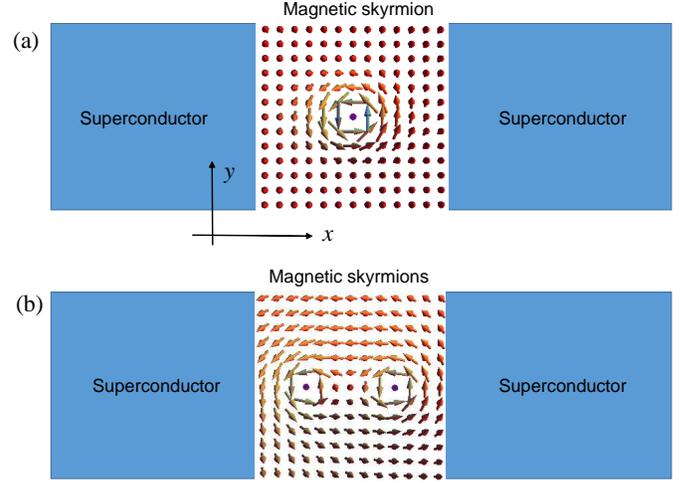}
}
\end{center}
\caption{(Color online) Schematic illustration of the superconductor/magnet/superconductor junction with (a) a single skyrmion and (b) two skyrmions. This setup may also be viewed as a simplified model for a lateral junction with superconducting electrodes deposited on top of a skyrmion thin film.}
\label{fig1}
\end{figure}

We consider a 2D superconductor / magnetic skyrmion / superconductor junction as shown in Fig. \ref{fig1}. This geometry is also expected to approximate well a planar junction geometry with two separated superconducting electrodes deposited on top of a thin magnetic film containing skyrmions. By assuming that the proximity effect is weak, we utilize the linearized Usadel equation:\cite{Usadel,Ivanov}
\begin{eqnarray}
D\nabla ^2 f_s  - 2\omega _n f_s  - 2i{\bf{f}}_t  \cdot {\bf{h}} = 0, \\ 
 D\nabla ^2 {\bf{f}}_t  - 2\omega _n {\bf{f}}_t  - 2if_s {\bf{h}} = 0.
\end{eqnarray}
Here, $D$ and $\omega _n$ are the diffusion constant in the magnet and Matsubara frequency, respectively. $f_s$ is the singlet anomalous Green's function while ${\bf{f}}_t$ represents the triplet anomalous Green's functions. ${\bf{h}} $ is the exchange field representing a magnetic structure with two skyrmions:\cite{Belavin}
\begin{align}
{\bf h} &= \frac{h}{{1 + {{\left| u \right|}^2}}}[2\text{Re} (u){\bf \hat{x}} +  2\text{Im} (u){\bf \hat{y}} + (1 - |u|^2){\bf \hat{z}}],\notag\\
u &= \frac{{i\lambda }}{{x - {x_c} - i(y - {y_c})}} + \frac{{i\lambda '}}{{x - x{'_c} - i(y - y{'_c})}}.
\end{align}
Here, $(x_c, y_c)$ and $(x'_c, y'_c)$ determine the centers of the two skyrmions. $\lambda$ and $\lambda'$ are the characteristic sizes of the skyrmions. 
The signs of $\lambda$ and $\lambda'$ determine the helicities of the skyrmions. $h$ is the magnitude of the exchange field. By setting $\lambda'=0$, the above exchange field represents a single skyrmion texture.
We consider the magnetic region in $- L/2 \le x,y \le L/2$. The interfaces are located at $x=\pm L/2$. 

The boundary condition at $x=-L/2$ reads \cite{Kupriyanov}
\begin{eqnarray}
- {\gamma _B}\xi \frac{{\partial {f_s}}}{{\partial x}} + {G_s}{f_s} = {F_s},\quad  - {\gamma _B}\xi \frac{{\partial {f_i}}}{{\partial x}} + {G_s}{f_i} = 0
\end{eqnarray}
where ${f_i}$ $(i=1,2,3)$ is the components of ${\bf{f}}_t$, and $G_s$ and $F_s$ are bulk Green's functions in the superconductor given by
\begin{eqnarray}
G_s  = \frac{{\omega _n }}{{\sqrt {\omega _n^2  + \Delta ^2 } }},\quad F_s  = \frac{{\Delta \exp \left( { - i\varphi /2} \right)}}{{\sqrt {\omega _n^2  + \Delta ^2 } }}.
\end{eqnarray}
Here, $\gamma _B$ describes the interface barrier strength, $\xi$ is the superconducting coherence length, $\Delta$ is the gap function, and $\varphi$ is the phase difference between the superconductors. The boudary condition at $x=L/2$ is given by changing the signs of the derivative and $\varphi$ in the above boundary condition. 
The boundary condition at $y=\pm L/2$ reads $\frac{{\partial {f_{\alpha}}}}{{\partial y}} = 0$ with $\alpha=s, 1,2,3$. The Josephson current is calculated as 
\begin{eqnarray}
\frac{{e{I_x}R}}{{2\pi {T_C}}} =  - \frac{T}{{{T_C}}}\sum\limits_{n \ge 0} {{\mathop{\rm Im}\nolimits} \left( {f_s^*{\partial _x}f_s - f_i^*{\partial _x}f_i} \right)}
\end{eqnarray}
with the (transition) temperature $T (T_C)$ and resistance of the magnet per length $R$. We define the total current as ${I_X} = \frac{1}{\xi }\int_{ - L/2}^{L/2} {{I_x}dy}$ at $x=-L/2$. The critical current and that including the sign of the current are denoted by ${I_{XC}}$ and $I{'_{XC}}$, respectively: ${I_{XC}} = \left| I{'_{XC}} \right|$. 
Below, we fix the parameters as $\gamma_B=10$, $T/T_C=0.9$ and $h/\Delta_0=1.5$ where $\Delta_0$ denotes the gap energy at zero temperature. Calculation of the Josephson current requires a solution of the 2D Usadel equation. We have solved the Usadel equation numerically by using an iterative method.

\begin{figure}[tbp]
\begin{center}
\scalebox{0.8}{
\includegraphics[width=9.50cm,clip]{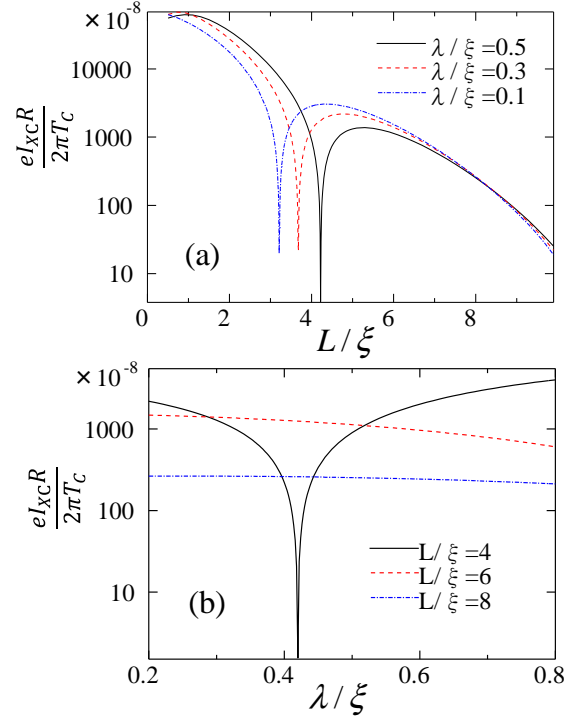}
}
\end{center}
\caption{(Color online) (a) The critical current as a function of the length of the magnetic region $L$ for several sizes of the skyrmion $\lambda$. (b) The critical current as a function of $\lambda$ for several $L$. We set $x_c=y_c=0$ and $\lambda'=0$.}
\label{fig2}
\end{figure}

\begin{figure}[tbp]
\begin{center}
\scalebox{0.8}{
\includegraphics[width=10.5cm,clip]{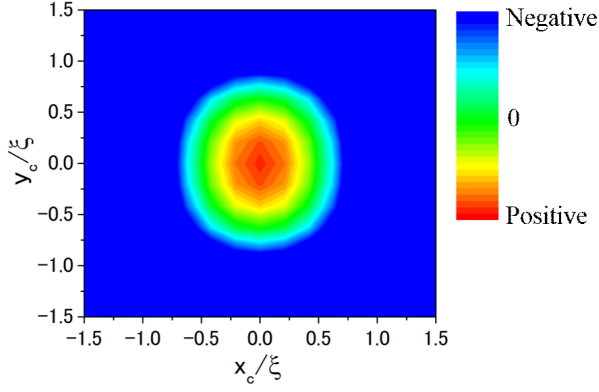}}
\end{center}
\caption{(Color online) The critical current including the sign of the current $\frac{{e{I{'_{XC}}}R}}{{2\pi {T_C}}}$ as a function of the position of the skyrmion $x_c$ and $y_c$ for $\lambda/\xi=0.5$, $\lambda'=0$, and $L/\xi=4.2$. This shows where the skyrmion should be located in the junction to induce a $0-\pi$ transition (the green circle). }
\label{fig3}
\end{figure}

\begin{figure}[tbp]
\begin{center}
\scalebox{0.8}{
\includegraphics[width=9.50cm,clip]{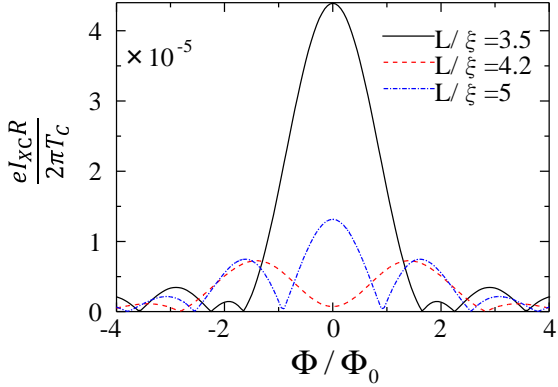}}
\end{center}
\caption{(Color online) The critical current as a function of the magnetic flux for several $L$. We set  $x_c=y_c=0$, $\lambda/\xi=0.5$ and $\lambda'=0$.}
\label{fig5}
\end{figure}

\begin{figure}[tbp]
\begin{center}
\scalebox{0.8}{
\includegraphics[width=8.5cm,clip]{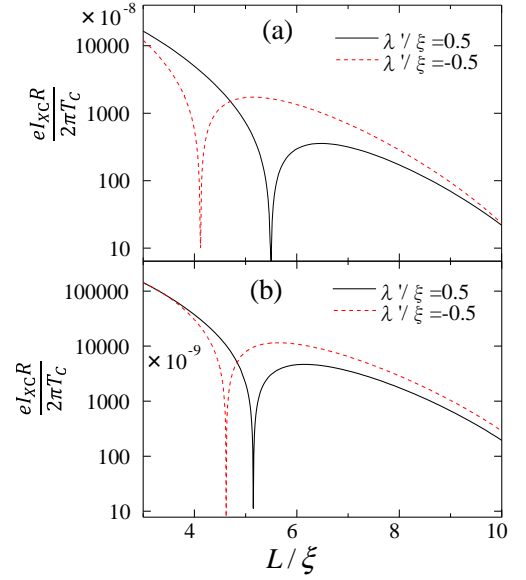}}
\end{center}
\caption{(Color online) The critical current as a function of the length of the magnetic region $L$ for $\lambda/\xi=0.5$ and either equal $(\lambda'/\xi=0.5)$ or opposite $(\lambda'/\xi=-0.5$) helicity of the two skyrmions. (a) $x_c/\xi=0.5$, $x'_c/\xi=-0.5$ and $y_c=y'_c=0$. (b) $x_c/\xi=1$, $x'_c/\xi=-1$ and $y_c=y'_c=0$.}
\label{fig4}
\end{figure}

We begin by considering junctions with a single skyrmion ($\lambda'=0$) as shown in Fig. \ref{fig1}(a). In Fig. \ref{fig2}(a), we show the critical current as a function of the length of the magnetic region $L$ for several sizes of the skyrmion $\lambda$. The skyrmion is assumed to be positioned in the center of the junction, $x_c=y_c=0$. We find a 0-$\pi$ transition as a function of $L$. It is also seen that the transition point can be controlled by altering the size of the skyrmion, $\lambda$.  In Fig. \ref{fig2}(b), we show the critical current as a function of $\lambda$ for several $L$ and $x_c=y_c=0$. For $L/\xi=4$, the 0-$\pi$ transition occurs around $\lambda/\xi=0.42$. 
These results indicate that the 0-$\pi$ transition is tunable by changing magnetic field and possibly also by applying an electric field/gate voltage since this breaks inversion symmetry and hence can modify the Dzyaloshinskii-Moriya interaction which in turn changes the size of the skyrmion.
The tunable size of skyrmions in helimagnetic alloys via spin-orbit coupling has been also experimentally verified in Ref. \cite{shibata_natnan_13}, indicating that the 0-$\pi$ transition predicted here can be manipulated via changing the skyrmion size according this route.

The next aspect we consider is how the skyrmion position influences the supercurrent response of the system. A unique feature of skyrmions is that the ultralow current density ($\sim 10^2$ A/cm$^{2}$) can induce their translational and/or rotational motions, which is typically 5 orders of magnitude smaller than the required density in conventional domain wall ferromagnets. This has been experimentally demonstrated in the helimagnet MnSi \cite{Jonietz} and FeGe \cite{yu_ncom_12}. Motivated by this, in Fig. \ref{fig3}, we show the critical current including the sign of the current as a function of the position of the skyrmion $x_c$ and $y_c$ for $\lambda/\xi=0.5$ and $L/\xi=4.2$. It is found that a 0-$\pi$ transition occurs by changing the position of the skyrmion. Since the position of the skyrmion can be manipulated by current or temperature gradient, this offers a way to control the quantum ground state of the system.

We also consider the critical current as a function of the magnetic flux threading through the magnetic region. 
We considier uniform magnetic field along the $z$-axis and include the vector potential of the form ${\bf{A}} = B( - y,0,0)$ by the substitution $\nabla  \to \nabla  + i\frac{2e}{\hbar }{\bf{A}}$ in the equations.
Figure \ref{fig5} exhibits the critical current as a function of the magnetic flux $\Phi$ for several $L$ with $\Phi  = B{L^2}$ and ${\Phi _0} = h/2e$. We find conventional Fraunhofer diffraction patterns for $L/\xi=3.5$ and 5. However, for $L/\xi=4.2$ near the 0-$\pi$ transition point as shown in Fig. 2(a), a minimum appears at $\Phi=0$ in the Fraunhofer pattern. This can be understood as follows. Due to the skyrmion magnetization texture varying along the $y$-axis, the present junction may be regarded as a parallel circuit of 0 and $\pi$ junctions. In such a circuit of 0 and $\pi$ junctions, a local minimum at $\Phi=0$ can appear due to the cancellation of the Josephson currents from the 0 and $\pi$ segments \cite{Kemmler}. In this way, the Fraunhofer pattern in our setup can display a local minimum at zero flux near the 0-$\pi$ transition points.

Now, let us consider junctions with two skyrmions (see Fig. \ref{fig1}(b)) and focus on the effect of the helicities. The presence of multiple skyrmions in the Josephson junction is particularly relevant in light of the experimental demonstration of multiple skyrmion configuration featuring skyrmions with both types of helicities \cite{yu_pnas_12,shibata_natnan_13}. The helicity was shown to be reversible via an external field of order $\sim 100$ mT \cite{yu_pnas_12}. Figure \ref{fig4} shows the critical current as a function of the length of the magnetic region $L$ for $\lambda/\xi=0.5$, considering both equal $(\lambda'/\xi=0.5)$ and opposite $(\lambda'/\xi=-0.5$) helicities of the two skyrmions. In Fig. \ref{fig4} (a), the positions of the skyrmions are set as $x_c/\xi=0.5$, $x'_c/\xi=-0.5$ and $y_c=y'_c=0$. Remarkably, we see that the 0-$\pi$ transition point depends on the helicity, which means that the reversal of the helicity can induce a 0-$\pi$ transition in itself. Moreover, this effect opens the possibility for a \textit{helicity-transistor effect} for the supercurrent: close to a 0-$\pi$ transition point for one helicity configuration, changing the relative helicity \textit{in situ} will result in an increase of the critical current density of several orders of magnitude as seen in Fig. \ref{fig4}. In Fig. \ref{fig4} (b), the distance of the skyrmions is set to be longer as $x_c/\xi=1$, $x'_c/\xi=-1$ and $y_c=y'_c=0$. It is found that the 0-$\pi$ transition points corresponding to the two helicity configurations become closer compared to Fig. \ref{fig4} (a). When two skyrmions are sufficiently separeted (e.g., for very large $x_c$), we can regard the two skyrmions as independent of each other and hence the effect of the helicity becomes negligible. 
Note that since the helicity changes under mirror operation with respect to $xy$ plane, the results remain the same under the reversal of all the helicities of the skyrmions: the results depend on the \textit{relative} sign of the helicities, and for a single skyrmion, the results do not depend on the helicity.

For the chiral magnet MnSi, the material parameters are estimated as $h \sim 1$ eV and $\lambda \sim 10$nm \cite{Zang}. 
The change of the exchange field $h$ will shift the 0-$\pi$ transiton point. 
Here, we have considered junctions with a single and two skyrmions. Skyrmions can also form a hexagonal lattice, and the application of our work to such a skyrmion configuration would be also informative. It could also be of interest to consider skyrmion tubes lined up along the junction direction, which should be experimentally feasible in layered thin-film structures. Moreover, the unusually low threshold for current-induced skyrmion motion would be very interesting to investigate in the context of supercurrent-induced magnetization dynamics and spin-transfer torques \cite{STT}. 
We leave these issues for future explorations.

In summary, we have investigated the Josephson effect in superconductor/magnetic skyrmion/superconductor junction.
It is found that the degrees of freedom associated with the skyrmions (size, position, and helicity), which recently have been demonstrated experimentally to be tunable via different routes, lead to a new dynamical way to control the 0-$\pi$ transition, offering the tantalizing prospect of a helicity-transistor for supercurrents.
It is also shown that the Fraunhofer pattern can exhibit a local minimum at zero flux as a consequence of the skyrmion magnetization texture.

\textit{Acknowledgments}. The authors thank S. Murakami, R. Takashima, A. Petrovic, and M. Ehrnstrom for helpful discussions. T.Y. was supported by Grant-in-Aid for Young Scientists (B) (No. 23740236), the "Topological Quantum Phenomena" (No. 25103709) Grant-in Aid for Scientific Research on Innovative Areas from the Ministry of Education, Culture, Sports, Science and Technology (MEXT) of Japan. J.L. acknowledges support from the Outstanding
Academic Fellows programme at NTNU, the COST
Action MP-1201' Novel Functionalities through Optimized
Confinement of Condensate and Fields', and the Norwegian
Research Council Grant No. 205591(FRINAT) and Grant No.
216700.


\begin{thebibliography}{99}

\bibitem{Buzdinrev} A. I. Buzdin, Rev. Mod. Phys. \textbf{77}, 935 (2005).

\bibitem{Bergeretrev} F. S. Bergeret, A. F. Volkov, and K. B. Efetov, Rev. Mod. Phys. \textbf{77}, 1321 (2005).


\bibitem{linder_nphys_15} J. Linder and J. W. A. Robinson, Nature Physics \textbf{11}, 307 (2015).


\bibitem{Bulaevskii} L. N. Bulaevskii, V. V. Kuzii, and A. A. Sobyanin, JETP
Lett. \textbf{25}, 290 (1977).

\bibitem{Buzdin} A. I. Buzdin, L. N. Bulaevskii, and S. V. Panjukov, JETP
Lett. \textbf{35}, 178 (1982).

\bibitem{Buzdin2} A. I. Buzdin, B. Bujicic, and B. M. Yu. Kupriyanov, Sov.
Phys. JETP \textbf{74}, 124 (1992).

\bibitem{Golubov} A. A. Golubov, M. Yu. Kupriyanov, and E. llichev, Rev. Mod. Phys. \textbf{76}, 411 (2004).


\bibitem{Ryazanov} V. V. Ryazanov, V. A. Oboznov, A. Yu. Rusanov, A. V.
Veretennikov, A. A. Golubov, and J. Aarts, Phys. Rev. Lett. \textbf{86},
2427 (2001).

\bibitem{Kontos} T. Kontos, M. Aprili, J. Lesueur, F. Genet, B. Stephanidis,
and R. Boursier Phys. Rev. Lett. \textbf{89}, 137007 (2002).


\bibitem{eschrig_07} M. Eschrig, T. L\"ofwander, T. Champel, J. C. Cuevas, J. Kopu, and G. Sch\"on, J. Low Temp. Phys. \textbf{147}, 457 (2007).


\bibitem{Bergeret3} F. S. Bergeret, A. F. Volkov, and K. B. Efetov, Phys. Rev. Lett. \textbf{86}, 4096 (2001); Phys. Rev. B \textbf{64}, 134506 (2001).

\bibitem{Keizer} R. S. Keizer, S. T. B. Goennenwein, T. M. Klapwijk, G. Miao, G. Xiao, and A. Gupta, 
Nature (London) \textbf{439}, 825 (2006).

\bibitem{Khaire} T. S. Khaire, M. A. Khasawneh, W. P. Pratt, Jr., and N. O. Birge, Phys. Rev. Lett. \textbf{104}, 137002 (2010).

\bibitem{Robinson} J. W. A. Robinson, J. D. S. Witt, M. G. Blamire, Science \textbf{329}, 59 (2010).

\bibitem{Eschrig} M. Eschrig, Physics Today \textbf{64}, 43 (2011). 


\bibitem{giazotto_prb_08} F. Giazotto and F. Taddei, Phys. Rev. B \textbf{77}, 132501 (2008).

\bibitem{romeo_prl_13} F. Romeo and R. Citro, Phys. Rev. Lett. \textbf{111}, 226801 (2013). 

\bibitem{trif_prl_13} M. Trif and Y. Tserkovnyak, Phys. Rev. Lett. \textbf{111}, 087602 (2013).

\bibitem{machon_prl_13} P. Machon, M. Eschrig, and W. Belzig, Phys. Rev. Lett. \textbf{110}, 047002 (2013). 

\bibitem{linder_prb_14} J. Linder and K. Halterman, Phys. Rev. B \textbf{90}, 104502 (2014). 


\bibitem{Martin} I. Martin and A. F. Morpurgo, Phys. Rev. B \textbf{85}, 144505 (2012).

\bibitem{Lu} Y.-M. Lu and Z. Wang, Phys. Rev. Lett. \textbf{110}, 096403 (2013).

\bibitem{Nakosai} S. Nakosai, Y. Tanaka, and N. Nagaosa, Phys. Rev. B \textbf{88}, 180503(R) (2013).


\bibitem{Rossler} U. K. R\"ossler, A. N. Bogdanov, and C. Pfleiderer, Nature (London) \textbf{442}, 797 (2006).

\bibitem{Muhlbauer} S. M\"uhlbauer, B. Binz, F. Jonietz, C. Pfleiderer, A. Rosch, A. Neubauer, R. Georgii, and P. B\"oni, Science \textbf{323}, 915 (2009).

\bibitem{Nagaosa2} N. Nagaosa and Y. Tokura, Nat. Nanotechnol. \textbf{8}, 899 (2013).

\bibitem{Lee} M. Lee, W. Kang, Y. Onose, Y. Tokura, and N. P. Ong, Phys. Rev. Lett. \textbf{102}, 186601 (2009).

\bibitem{Neubauer} A. Neubauer, C. Pfleiderer, B. Binz, A. Rosch, R. Ritz, P. G. Niklowitz, and P. B\"{o}ni, Phys. Rev. Lett. \textbf{102}, 186602 (2009).

\bibitem{Zang} J. Zang, M. Mostovoy, J. H. Han, and N. Nagaosa, Phys. Rev. Lett. \textbf{107}, 136804 (2011).

\bibitem{Jonietz} F. Jonietz, S. Muhlbauer, C. Pfleiderer, A. Neubauer, W. Munzer, A. Bauer, T. Adams, R. Georgii, P. Boni, R. A. Duine, K. Everschor, M. Garst, and A. Rosch, Science \textbf{330}, 1648 (2010).

\bibitem{Fert} A. Fert, V. Cros, and J. Sampaio, Nat. Nanotechnol. \textbf{8}, 152 (2013).

\bibitem{Iwasaki} J. Iwasaki, M. Mochizuki, and N. Nagaosa, Nat. Commun. \textbf{4}, 1463 (2013).

\bibitem{Iwasaki2} J. Iwasaki, M. Mochizuki, and N. Nagaosa, Nat. Nanotechnol. \textbf{8}, 742 (2013).

\bibitem{Troncoso} R. E. Troncoso and A. S. N\'{u}\~{n}ez, Phys. Rev. B \textbf{89}, 224403 (2014).

\bibitem{Kong} L. Kong and J. Zang, Phys. Rev. Lett. \textbf{111}, 067203 (2013).

\bibitem{Mochizuki} M. Mochizuki, X. Z. Yu, S. Seki, N. Kanazawa, W. Koshibae, J. Zang, M. Mostovoy, Y. Tokura, and N. Nagaosa, Nat. Mater. \textbf{13}, 241 (2014).

\bibitem{Lin} S. Z. Lin, C.D. Batista, C. Reichhardt, and A. Saxena, Phys. Rev. Lett. \textbf{112}, 187203 (2014).

\bibitem{Kovalev} A. A. Kovalev, Phys. Rev. B \textbf{89}, 241101(R) (2014).

\bibitem{yu_pnas_12} X. Yu, M. Mostovoy, Y. Tokunaga, W. Zhang, K. Kimoto, Y. Matsui, Y. Kaneko, N. Nagaosa, and Y. Tokura, Proc. Natl. Acad. Sci. USA \textbf{109}, 8856 (2012).

\bibitem{shibata_natnan_13} K. Shibata, X. Z. Yu, T. Hara, D. Morikawa, N. Kanazawa, K. Kimoto, S. Ishiwata, Y. Matsui, and Y. Tokura, Nature Nanotechnology \textbf{8}, 723 (2013).


\bibitem{yu_ncom_12} X. Z. Yu, N. Kanazawa, W. Z. Zhang, T. Nagai, T. Hara, K. Kimoto, Y. Matsui, Y. Onose, and Y. Tokura, Nat. Commun. \textbf{3}, 988 (2012).


\bibitem{Usadel} K. D. Usadel, Phys. Rev. Lett. \textbf{25}, 507 (1970).

\bibitem{Ivanov} D. A. Ivanov and Ya. V. Fominov, Phys. Rev. B \textbf{73}, 214524 (2006).

\bibitem{Belavin} A. A. Belavin and A. M. Polyakov, Pis'ma Zh. Eksp. Teor. Fiz. \textbf{22}, 503 (1975) [JETP Lett. \textbf{22}, 245 (1975)].

\bibitem{Kupriyanov} M. Yu. Kupriyanov and V. F. Lukichev, Zh. Eksp. Teor. Fiz. \textbf{94}, 139 (1988) [Sov. Phys. JETP \textbf{67}, 1163 (1988)].

\bibitem{Kemmler} M. Kemmler, M. Weides, M. Weiler, M. Opel, S. T. B. Goennenwein, A. S. Vasenko, A. A. Golubov, H. Kohlstedt, D. Koelle, R. Kleiner, and E. Goldobin, Phys. Rev. B \textbf{81}, 054522 (2010); M. Alidoust, G. Sewell, and J. Linder, Phys. Rev. Lett. \textbf{108}, 037001 (2012).

\bibitem{STT} X. Waintal and P. W. Brouwer, Phys. Rev. B \textbf{65}, 054407 (2002); E. Zhao and J. A. Sauls, Phys. Rev. B \textbf{78}, 174511 (2008); F. Konschelle and A. Buzdin, Phys. Rev. Lett. \textbf{102}, 017001 (2009); J. Linder and T. Yokoyama, Phys. Rev. B \textbf{83}, 012501 (2011); J. Linder, A. Brataas, Z. Shomali, and M. Zareyan, Phys. Rev. Lett. \textbf{109}, 237206 (2012); B. Baek, W. H. Rippard, M. R. Pufall, S. P. Benz, S. E. Russek, H. Rogalla, and P. D. Dresselhaus, Phys. Rev. Applied \textbf{3}, 011001 (2015).




\end{thebibliography}
\end{document}